\documentclass{emulateapj}

\newcommand{\rxte}{{\em RXTE\ }}
\newcommand{\rxteno}{{\em RXTE}}
\newcommand{\exosat}{{\em EXOSAT\ }}

\newcommand{\Hzs}{Hz s$^{-1}$} 
\begin{document}
\title{Discovery of a Transition to Global Spin-up in EXO 2030+375}
\author{Colleen A. Wilson}
\affil{XD 12 Space Science Branch, National Space Science and 
Technology Center, 320 Sparkman Drive, Huntsville, AL 35805}
\email{colleen.wilson@nasa.gov}
\author{Juan Fabregat}
\affil{Dept d'Astronomia i Astrofisica, Universitat de Valencia, 46100 
Burjassot, Valencia, Spain}
\email{juan@pleione.uv.es}
\author{Wayne Coburn}
\affil{Space Sciences Laboratory, University of California at Berkeley, Grizzly
Peak at Centennial Drive, Berkeley, CA 94720-7450}
\email{wcoburn@ssl.berkeley.edu}
\begin{abstract}
EXO 2030+375, a 42-second transient X-ray pulsar with a Be star
companion, has been observed to undergo an outburst at nearly every 
periastron passage for the last 13.5 years. From 1994 through 2002, the
global trend in the pulsar spin frequency was spin-down. Using \rxte
data from 2003 September, we have observed a transition to global
spin-up in EXO 2030+375. Although the spin frequency observations are 
sparse, the relative spin-up between 2002 June and 2003 September 
observations, along with an overall brightening of the outbursts since 
mid 2002 observed with the \rxte ASM, accompanied by an increase in 
density of the Be disk, indicated by infrared magnitudes, suggest that the
pattern observed with BATSE of a roughly constant spin frequency, followed by 
spin-up, followed by spin-down is repeating. If so this pattern has 
approximately an 11 year period, similar to the $15 \pm 3$ year period derived 
by \citet{Wilson02} for the precession period of a one-armed oscillation in
the Be disk. If this pattern is indeed repeating, we predict a transition
from spin-up to spin-down in 2005.
\end{abstract}
\section{Introduction}
\subsection{Be/X-ray binaries}
Be/X-ray transients, the most common type of accreting X-ray pulsar system,
consist of a pulsar and a Be (or Oe) star, a main sequence star of spectral
type B (or O) that shows Balmer emission lines 
\citep[See e.g.,][for a review.]{Porter03} The line emission is believed to be 
associated with circumstellar material shed by the Be star into its equatorial 
plane by an unknown mass loss process, thought to be related to the rapid 
rotation of the Be star, typically near 70\% of the critical break-up velocity 
\citep{Porter96}. Near the Be star, the equatorial outflow probably forms a 
quasi-Keplerian disk \citep{Quirrenbach97,Hanuschik96}.

Be/X-ray binaries typically show three types of behavior: (a) giant 
outbursts (or type II), characterized by high luminosities and high spin-up 
rates  (i.e., a significant increase in pulse frequency), (b) normal 
outbursts (or type I), characterized  by lower luminosities, low spin-up rates
(if any), and repeated occurrence at the orbital period, and (c) quiescence,
where the accretion is partially or completely halted
\citep{Stella86,Motch91,Bildsten97}.

For isolated Be stars, variations in the infrared bands are believed to be good
indicators of the size of the Be star's disk. However, when the Be star is in a
binary system with a neutron star, the Be disk is truncated at a resonance 
radius by tidal forces from the orbit of the neutron star \citep{Okazaki01}. In
these systems, as the disk cannot easily change size because of the truncation 
radius, changes in mass loss from the Be star produce changes in the disk 
density, which can even become optically thick at infrared wavelengths 
\citep[see, e.g.,][]{Neg01a,Mir01}. 

In  most systems there is no clear correlation between X-ray outbursts and 
optical activity within single outbursts. The X-ray activity however, follows 
the long-term optical activity cycle of the Be star, in the sense that no 
outbursts occur in periods where optical indicators of the Be star disk, such as
$H\alpha$ emission, have disappeared. Periods of X-ray quiescence when the Be 
disk is present are also observed; these may be due to the truncation of the Be
star disk well within the neutron star orbit \citep{Neg01a, Negueruela01b}.

\subsection{EXO 2030+375}
EXO 2030+375 is a 42-second transient accreting X-ray pulsar with a B0 Ve
star companion \citep{Motch87,Janot88,Coe88} discovered during a giant outburst
in 1985 \citep{Parmar89}. The most extensive observations of EXO 2030+375 were 
from nearly continuous monitoring of pulse frequency and pulsed flux with the 
Burst and Transient Source Experiment (BATSE) on the {\em Compton Gamma Ray 
Observatory (CGRO)} from 1991 April until 2000 June \citep{Wilson02}. 
We \citep{Wilson02} found from BATSE and {\em Rossi X-ray Timing Explorer 
(RXTE)} data that EXO 2030+375 appeared to have undergone an outburst near most
likely every periastron passage for the last 13 years. Our BATSE observations
revealed that EXO 2030+375's spin frequency remained roughly constant for about a
year, followed by 2 years of spin-up and 6.5 years of spin-down. 

The long baseline of X-ray measurements allowed us to make detailed comparisons 
with optical H$\alpha$ and infrared (IR) measurements, which led to the following
interpretation \citep{Wilson02}: around MJD 49000, a major reconfiguration occurred
in the Be star's disk, resulting in a much lower density as indicated by the 
fainter K-band magnitudes. The lower density meant less matter was available to
be accreted, and as a consequence, the X-ray flux dropped and the spin-up of the
neutron star ended. At the same time or shortly after, a density perturbation 
developed in the disk and started to precess without interacting with the 
neutron star's orbit. Around MJD 50000 the density perturbation interacted with 
the neutron star's orbit at a phase corresponding to about 2.5 days before 
periastron, producing X-ray outbursts that peak at that time. The density 
perturbation precessed in a prograde direction around the Be disk, changing the 
orbital phase of the outburst peaks. At about MJD 50700, the density perturbation
lost contact with the neutron star's orbit in a position symmetrical with respect
to periastron, at about 2.5 days after periastron. This ended the fast migration 
of the outburst peaks. \citet{Wilson02} interpreted the trend in the orbital 
phases from MJD 50000 to 50600 (1995 October-1997 June) in terms of a beat 
frequency between the orbital period and a perturbation period, assuming that the
perturbation period was longer than the orbital period, as is typical for 
one-armed oscillations (density perturbations), and obtained a perturbation 
period of $15 \pm 3$ years.

\section{Observations and Analysis}

In 2002 June (MJD 52425-52446) and 2003 September (MJD 52894-52898), we observed
two outbursts of EXO 2030+375 with the \rxte Proportional Counter Array
(PCA)\footnote{See \url{http://heasarc.gsfc.nasa.gov/} for observation details}.
For each observation, we extracted a barycentered light curve using FTOOLS 
v5.3.1 for PCA Standard 1 data (0.125 sec time resolution, no energy 
resolution.) Arrival times were corrected using the orbital ephemeris of
\citet{Wilson02}. These data were then fitted with a model consisting of a 
constant background plus a sixth-order Fourier expansion in the pulse phase 
model, which consisted of a constant barycentric frequency ($\nu_0 = 23.9852$ 
and 23.9880 mHz for the 2002 and 2003 outbursts, respectively), generating a 
pulse profile for each observation. Phase offsets to the constant frequency
model were then generated by cross-correlating the 2-60 keV pulse profiles from
each observation with the template profile from 1996 July 4 used in 
\cite{Wilson02}.  These pulse phase measurements were combined with frequency
measurements from \exosat and phase measurements from BATSE and \rxte described
in \citet{Wilson02}. As done in \citet{Wilson02}, we fitted the all of the data
with a global orbit plus a third-order polynomial intrinsic pulse frequency model
for the \exosat-observed frequencies and a different quadratic pulse phase model
for each outburst in the BATSE or \rxte data, extending the fit of
\citet{Wilson02} by 2 additional \rxte outbursts and extending the
baseline spanned by observations to 18 years. The best fit orbital
parameters: $P_{\rm orb} = 46.0202(2)$ d, 
$T_{\rm peri} = {\rm JD}2451099.93(2)$, $a_{\rm X} \sin i = 238(2)$ lt-s,
$e = 0.416(1)$, $\omega = 210.8(4)\arcdeg$, with $\chi^2/{\rm dof} = 646.4/373$,
were consistent with those derived by \citet{Wilson02}. Within the 2003 
outburst, our model suggests at a 2 $\sigma$ level that spin-up is 
present, with an estimated spin-up rate of $(2.4 \pm 1.2) \times
10^{-13}$ \Hzs. 

Infrared photometry in the JHK bands was obtained as part of a
monitoring program of Be/X-ray binaries at the 1.5 m. Carlos S\'anchez
Telescope (TCS), located at the Teide Observatory in Tenerife, Spain.
The instruments used were the Continuously Variable Filter Photometer
(CVF) up to October 1999, and the Cain-II camera, equipped with a 
$256\times 256$ HgCdTe (NICMOS 3) detector ever since.  Instrumental CVF
and CAIN-II values were transformed respectively to the standard systems
defined by \citet{Alonso98} and  \citet{Hunt98}. The accuracy of
the standard JHK values is 0.01 magnitudes (CVF) and 0.03 magnitudes (CAIN-II).

\section{Results}

\begin{figure}
\plotone{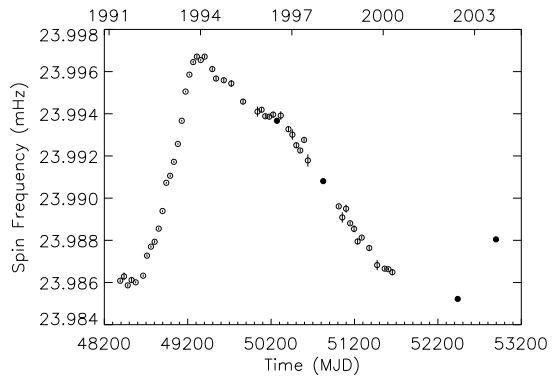}
\caption{Barycentered, orbit corrected, spin frequency measured 
with BATSE (open circles) and \rxte PCA (filled circles.) \rxte PCA 
measurements in 2003 September clearly indicate that EXO 2030+375 is 
again spinning up. \label{fig:freq}}
\end{figure}

Figure~\ref{fig:freq} shows the best fit EXO 2030+375 spin frequencies 
extracted for each BATSE and \rxte outburst from our model. Between the last 
outburst detected with BATSE in 2000 April and the \rxte PCA observed outburst 
in 2002 June, the average global spin-down rate was $(-1.9 \pm 0.2) \times 
10^{-14}$ \Hzs. This is less than half the average spin-down rate observed by 
\citet{Wilson02} with BATSE, suggesting that the rate of spin-down had slowed by
2002. 

From Figure~\ref{fig:freq}, one can clearly see that between 2002 June
and 2003 September, the global trend in EXO 2030+375 changed to spin-up with an
average rate of $(7.1 \pm 0.2) \times 10^{-14}$ \Hzs. This rate was about 40\%
of the average global spin-up rate observed from 1992 February to 1993 November
with BATSE \citep{Wilson02}, suggesting that EXO 2030+375 likely spent some of
the interval between the two \rxte observations in a roughly constant spin
frequency state similar to that observed with BATSE in 1991 through early 1992.
This possibility is intriguing and suggests that EXO 2030+375 may be repeating
the pattern of constant spin, followed by spin-up, followed by spin-down
observed with BATSE. To further check this, we compared average spin-up rates 
between BATSE outbursts spaced by about 460 days, the interval between the \rxte
observations. The average spin-up rate between the 1992 August and 1991 May
outbursts\footnote{\citep[See ][Table 1, for detailed times.]{Wilson02}} was 
$(6.2 \pm 0.2) \times 10^{-14}$ \Hzs\ and $(7.8 \pm 0.2) \times 10^{-14}$ \Hzs\
between the 1992 September-October and 1991 June-July outbursts, quite similar 
to what we observed with \rxteno. If the pattern is indeed repeating then it 
appears to have a period of about 11 years.

\begin{figure}
\plotone{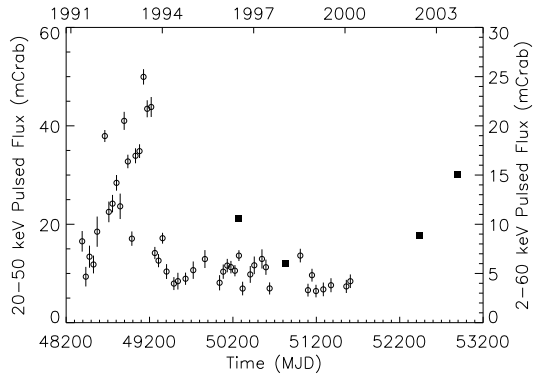}
\caption{Peak rms pulsed flux measured with BATSE (20-50 keV, 4-day averages, 
open circles, left-hand vertical scale) and with \rxte PCA (2-60 keV, filled squares,
right-hand vertical scale.) Error bars on the \rxte PCA measurements are smaller than the
symbols. The definitions used for 1 mCrab are $1 \times 10^{-11}$ ergs
cm$^{-2}$ s$^{-1}$ and 2.786 counts s$^{-1}$ PCU$^{-1}$ for BATSE and the
\rxte PCA, respectively. Details of how pulsed fluxes were measured are
given in \citet{Wilson02}.\label{fig:pflux}}
\end{figure}

\begin{figure}
\epsscale{1.25}
\plotone{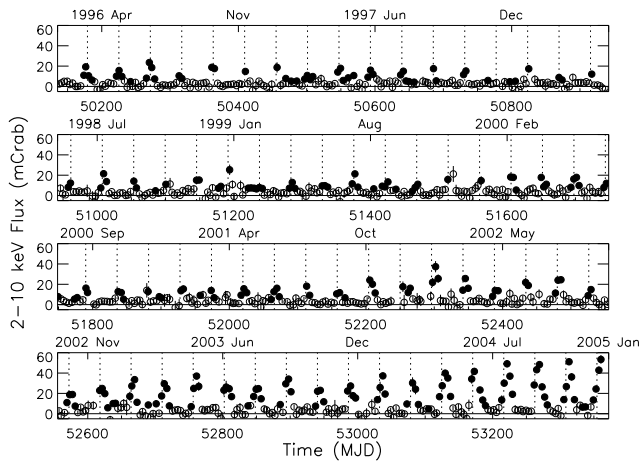}
\caption{EXO 2030+375 flux measured in 4-day averages with the \rxte ASM.
Dashed vertical lines denote predicted time of periastron passages. Filled
circles denote 3-$\sigma$ or better detections.
\label{fig:asm}}
\end{figure}

As the pulsar spun-up in the BATSE era, the peak pulsed flux of the 
outbursts increased, a trend which appears to be recurring in the \rxte data. 
Figure~\ref{fig:pflux} shows the peak pulsed flux measured with
BATSE and/or the \rxte PCA for each outburst computed as described in 
\citet{Wilson02}. Comparing the peak 2-60 keV pulsed fluxes measured with the 
\rxte PCA in Figure~\ref{fig:pflux} (filled squares) indicates that the pulsed 
flux has clearly increased in the 2003 observation and is the brightest observed
with \rxte to date. The BATSE and \rxte pulsed fluxes cannot easily be directly
compared since they are from different energy ranges and are computed using
different methods. Figure~\ref{fig:asm} shows 4-day average 2-10 keV total flux
measured with the \rxte All-Sky Monitor (ASM) from 1996 January to 2004 
December.  From 1996 through late 2001, the outbursts were very faint. In early
to mid 2002, the outbursts began to brighten and have continued to do so 
throughout 2004.  {\em INTEGRAL} also detected the increasing outburst flux 
\citep{Gotz04}.

\begin{figure}
\epsscale{1.0}
\plotone{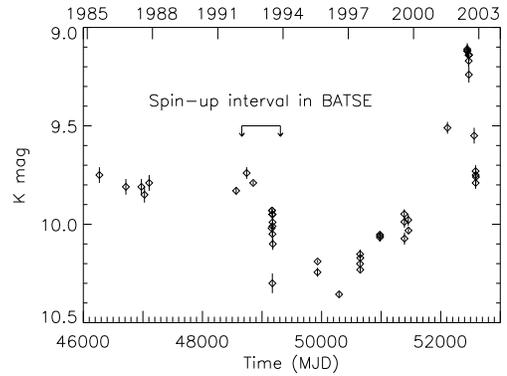}
\caption{K-band infrared magnitudes for EXO 2030+375. Recent observations
suggest that the Be disk has again brightened.\label{fig:kband}}
\end{figure}

Figure~\ref{fig:kband} shows the long term history of infrared K magnitudes.
At approximately the same time when the 
outburst flux dropped dramatically in the BATSE data, the K-band infrared flux 
from the companion also dropped, indicating a
decrease in density of the Be star's disk \citep{Wilson02}.  The last 
measurements shown in \citet{Wilson02} indicated that by late 1999, the
Be disk had brightened, i.e. increased in density, but not yet to the
level observed when EXO 2030+375 was previously spinning-up. We had no
observations from late 1999 until 2001 July when the K band magnitudes 
were brighter than when spin-up was observed with BATSE. In 2002
June-July, the Be disk was very bright. By 2002 November, it had
returned to the level observed during the bright BATSE outbursts.
Brightening of the X-ray outbursts as observed by the \rxte ASM appeared
to approximately coincide with the brightening of the Be disk,
indicating that the disk density had again increased. Over the long
term, the Be disk density appears to be driving the X-ray outburst
intensity.

\begin{figure}
\plotone{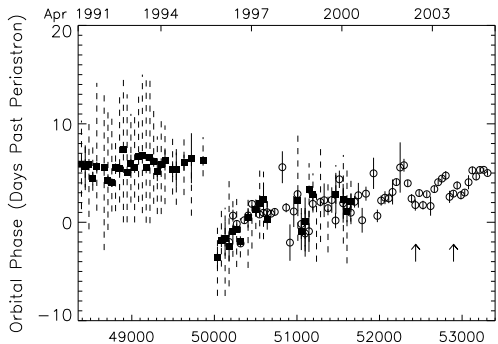}
\caption{Orbital phase of EXO 2030+375 outburst peaks versus time. Filled 
squares indicate the times of outburst peaks measured with BATSE. Open circles 
indicate the times of outburst peaks estimated from the \rxte ASM. Arrows indicate
the times of the 2002 and 2003 outbursts observed with the \rxte PCA.
\label{fig:orbph}}
\end{figure}

\citet{Wilson02} observed a sudden shift in the orbital phase of EXO 2030+375's
outbursts from about 6 days after periastron to about 4 days before periastron,
followed by a rapid recovery to peaking about 2.5 days after periastron,
interpreted by \citet{Wilson02} as a new stable orbital phase. However, we have 
continued to monitor the orbital phase of the outbursts using \rxte ASM outbursts
and have found that the orbital phase of the outbursts continued to gradually
change. As of 2004 November, the orbital phase of the outburst peak has reached 
about 5 days after periastron. Figure~\ref{fig:orbph} shows the orbital phase of 
EXO 2030+375's outburst peaks from 1991 April-2004 November. The BATSE outburst 
peaks are taken from data included in Figure~9 in \citet{Wilson02}. For each 
periastron passage where \rxte ASM data were available, we fitted a Gaussian to 
46.02 days of single dwell, 2-10 keV data, centered on the predicted periastron 
passage time, to determine the peak time of the outburst.  

\section{Conclusions}

Based on the 2002 and 2003 outbursts we observed with the \rxte PCA, the global
trend for EXO 2030+375 has reversed, changing from spin-down to spin-up. As this
change was occurring, \rxte ASM data indicated that the outbursts were brightening
and infrared data indicated that the Be disk density had again increased to at
or above its density when BATSE observed spin-up and bright outbursts. The
combination of these effects and the similar average spin-up rate between early 
BATSE outbursts spaced by 460 days (the spacing between the \rxte PCA
observations), suggests that the pattern of constant spin, followed by spin-up,
followed by spin-down, observed with BATSE, is repeating with an approximately 
11 year cycle. If this pattern continues, we predict that EXO 2030+375 will
transition to spin-down in 2005. 

The pattern we observe appears to be related to the density of the Be disk, but
it is unclear if it is related to the density perturbation that caused the shift
in outburst phase. Interestingly, the 11 year period is similar to the $15 \pm 
3$ year period derived by \citet{Wilson02} for the propagation period of a 
density perturbation around the Be disk.  However, this density perturbation was
used to explain the shift in orbital phase of the outburst peaks and the peak 
phase has not yet returned to 6 days after periastron as was observed with 
BATSE, suggesting that this cycle may not yet be complete. Evidence for 
quasi-periods in the H$\alpha$ line profiles (so called V/R variability) has 
been seen in many Be/X-ray binaries with periods ranging from a few weeks to 
years \citep{Negueruela98}. This V/R variability is believed to be due to 
density perturbations in the Be disk. Other Be/X-ray binaries (e.g., 4U 
0115+634) have undergone giant outbursts with near simultaneous asymmetry in the
H$\alpha$ profile, indicating a perturbation at the distance of the neutron star
orbit. However, EXO 2030+375 was the first object found to exhibit a sudden 
shift in outburst phase of normal outbursts that was likely correlated with a 
density perturbation in the Be disk \citep{Wilson02}. To date, the quasi-periods
shown in Be/X-ray transients are much shorter than those found in isolated Be 
stars, whose periods range from 2 years to decades. Perhaps EXO 2030+375 is a 
counter-example, a Be transient with a decade long quasi-period. 

\acknowledgements 
The Carlos S\'anchez Telescope is operated in the Spanish Observatorio
del Teide by the Instituto de Astrof{\'\i}sica de Canarias. \rxte ASM
results provided by the ASM/\rxte teams at MIT and at the \rxte SOF and GOF at 
NASA's GSFC.

\end{document}